\documentclass[12pt, aaspp4, flushrt, mathptmx]{aastex}
\usepackage{epsf}
\usepackage{verbatim}

\newcommand{\comments}[1]{}

\begin{document}
\sffamily

\title{Characteristic Length of Energy-Containing Structures at the Base of a Coronal Hole}

\author{Abramenko,V.I.$^1$,  Zank, G.P.$^2$, Dosch, A.$^2$, Yurchyshyn,
V.B.$^1$, Goode, P.R.$^1$, Ahn, K.$^1$, Cao, W.$^1$}

\affil{
$^1$ Big Bear Solar Observatory, 40386 N. Shore Lane, Big Bear City, CA
92314, USA\\
$^2$ CSPAR, University of Alabama in Huntsville, Huntsville, AL, USA
}

\begin{abstract}

An essential parameter for models of coronal heating and fast solar wind
acceleration that rely on the dissipation of MHD turbulence is the
characteristic energy-containing length $\lambda_{\bot}$ of the squared
velocity and magnetic field fluctuations ($u^2$ and $b^2$) transverse to the
mean magnetic field inside a coronal hole (CH) at the base of the corona. The
characteristic length scale defines directly the heating rate. We use a time
series analysis of solar granulation and magnetic field measurements inside two
CHs obtained with the New Solar Telescope (NST) at Big Bear Solar Observatory.
A data set for transverse
magnetic fields obtained with the Solar Optical Telescope/Spectro-Polarimeter
(SOT/SP) aboard {\it Hinode} spacecraft was utilized to analyze the squared
transverse magnetic field fluctuations $b_t^2$. Local correlation tracking
(LCT) was applied to derive the squared transverse velocity fluctuations $u^2$.
We find that for $u^2$-structures, Batchelor integral scale $\lambda$ varies
in a range of 1800 - 2100 km, whereas the correlation length $\varsigma$  and
the $e$-folding length $L$ vary between 660 and 1460 km. Structures for
$b_t^2$ yield $\lambda \approx 1600$~km, $\varsigma \approx 640$~km, and $L
\approx 620$~km. An averaged (over $\lambda, \varsigma$, and $L$) value of the
characteristic length of $u^2$-fluctuations is 1260$\pm$500~km, and that of
$b_t^2$ is 950$\pm$560~km. The characteristic length scale in the photosphere is
approximately 1.5-50 times smaller than that adopted in previous models
(3-30$\times$10$^3$~km). Our results provide a critical input parameter for
current models of coronal heating and should yield an improved understanding of
fast solar wind acceleration.

\end{abstract}


\section{Introduction}

Solar magneto-convection in the photosphere and beneath gives rise to a broad
spectrum of magneto-hydrodynamic (MHD) waves, which carry energy (both kinetic
and magnetic) from the base of the solar corona into its outer parts. There is
no doubt that the propagation, absorption, interaction and reflection of various
types of MHD waves significantly contribute to the acceleration of solar wind 
\citep[e.g.][to mention a few]{Matthaeus+1999,Oughton+1999,Thomas-Stanch-2000,Dmitruk+2001,Bogdan+2003,
Cranmer+vanBall-2005, Cranmer+2007,Verdini+2010,Zank+2012}. To account for
turbulent dissipation of MHD waves, a solution to the dissipative MHD equations
is required, i.e., the dissipative term should be incorporated. The dissipative
term, in turn, depends on a "free" parameter, namely, the characteristic
energy-containing length $\lambda_{\perp}$ of the dynamical structures
transverse to the mean magnetic field in a coronal hole (CH) at the base of the
corona, where the fast solar wind is emanated \citep[e.g.][]{Matthaeus+1999,Oughton+1999,
Dmitruk+2001,Cranmer+vanBall-2005, Zank+2012}. The turbulent dissipation rate,
responsible for heating and solar wind acceleration, is inversely proportional
to $\lambda_{\perp}$, see Eq (8) in \citet{Dmitruk+2001}, Eq. (56) in \citet
{Cranmer+vanBall-2005} and Eq. (16) in \citet{Zank+2012}.

The characteristic energy-containing length scale is an integral element
underlying the decomposition of MHD fluctuations into high- and low-frequency
components \citep{Matthaeus+1999, Zank+2012}. Indeed, within the existing models
that exploit the dissipation of low-frequency turbulence to heat the solar
corona, $\lambda_{\perp}$ plays perhaps the key role in determining efficiency
of heating. Rather surprisingly, almost nothing is known observationally about
this critical parameter. Currently, only a very rough estimate of
$\lambda_{\perp}$ was obtained based on the fact that the network spacing is 
about 3$\times$10$^4$ km \citep[or an order of magnitude less,][]
{Matthaeus+1999, Dmitruk+2001}. \citet{Cranmer+vanBall-2005} introduced 
parametrized $\lambda_{\perp}$, which could be translated into 3 Mm near the
base of the corona.  

The goal of this work is to provide a better estimate for the energy-containing
length scales using observed data on the transverse velocities and magnetic
fields at the coronal base of a CH. We will analyse energy-containing structures
of $u^2 = u_x^2 + u_y^2$ and $b_t^2$, where $u_x$, and $ u_y$ are the transverse
velocities, and $b_t$ is the transverse component of the magnetic field in the
photosphere inside a CH. Only for the purposes of comparison, the line-of-sight
(LOS) component of the magnetic field is also analysed.

\section{Data}

We utilize data from the 1.6 m clear aperture New Solar Telescope (NST,
\citet{Goode_BPs_2010}) operated at the Big Bear Solar Observatory. High time
cadence sequences of solar granulation images allowed us to calculate squared
transverse velocities at the photospheric level ($\tau_{500}=1$), whereas the
co-temporal and co-spatial near-infrared magnetograms from the NST provided us
with the LOS magnetic fields inside CHs. The squared transverse magnetic field
fluctuations $b_t^2$ analyzed here were obtained from the Spectro-Polarimeter
data at the Solar Optical Telescope (SOT/SP) aboard {\it Hinode} spacecraft
\citep{Kosugi_2007,tsuneta2008} acquired on 2007 March 10. 

Two CHs for which NST time series were obtained are shown in Figure 1. Both CHs
were observed at the times of their crossing the central meridian. The first CH
was observed on 2011 August 12, and it is referred to hereafter as CH
2011-08-12. It was observed for nearly 20 minutes with a field of
view (FOV, 56$''$ $\times$ 56$''$). The second CH was observed with a FOV of
31$''$ $\times$ 28$''$ during about 2 hours on 2012 June 4 and is referred to
hereafter as CH 2012-06-04. 
%#####################################################################
\begin{figure}[!ht]
\centerline{
\epsfxsize=3.5truein \epsffile{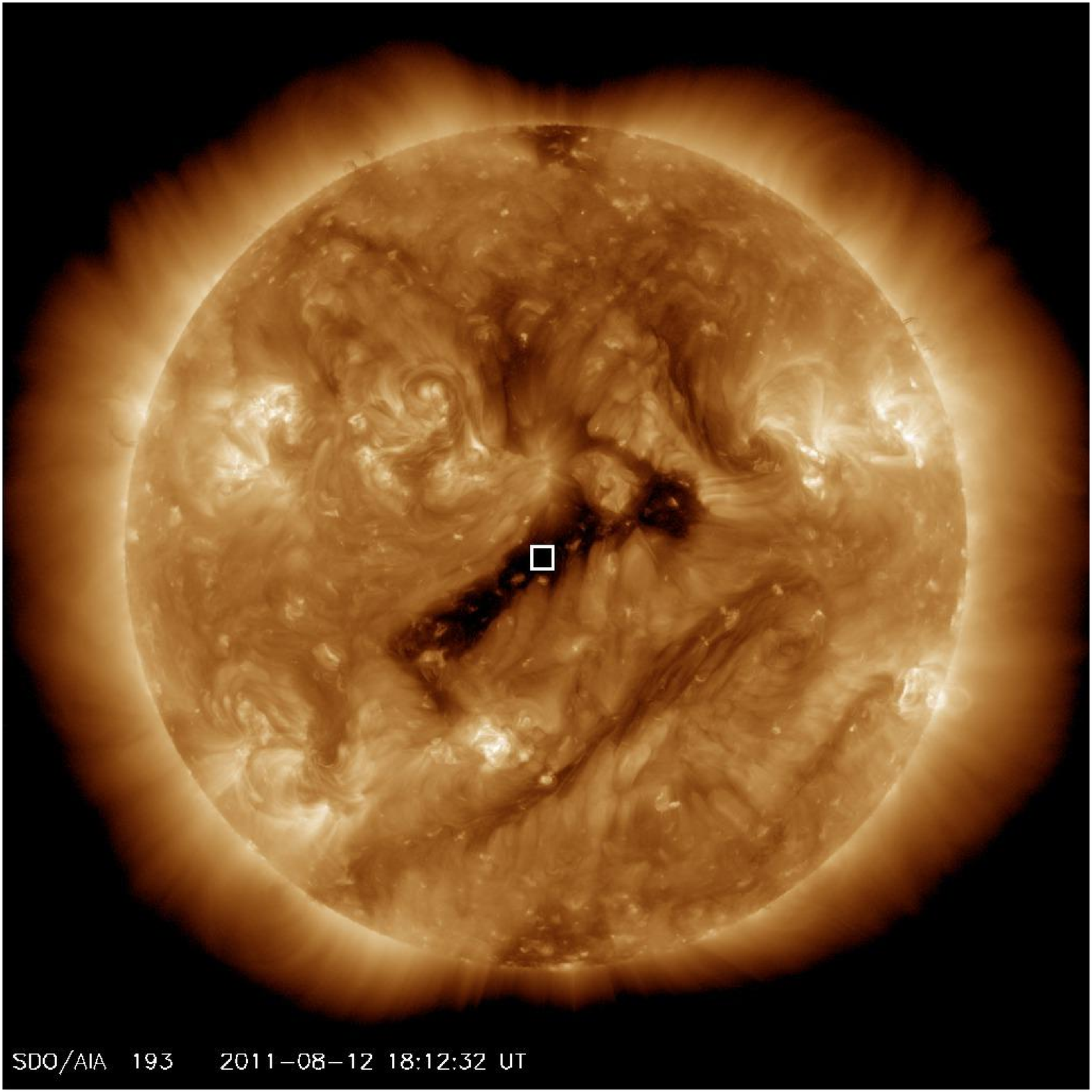}
\epsfxsize=3.5truein \epsffile{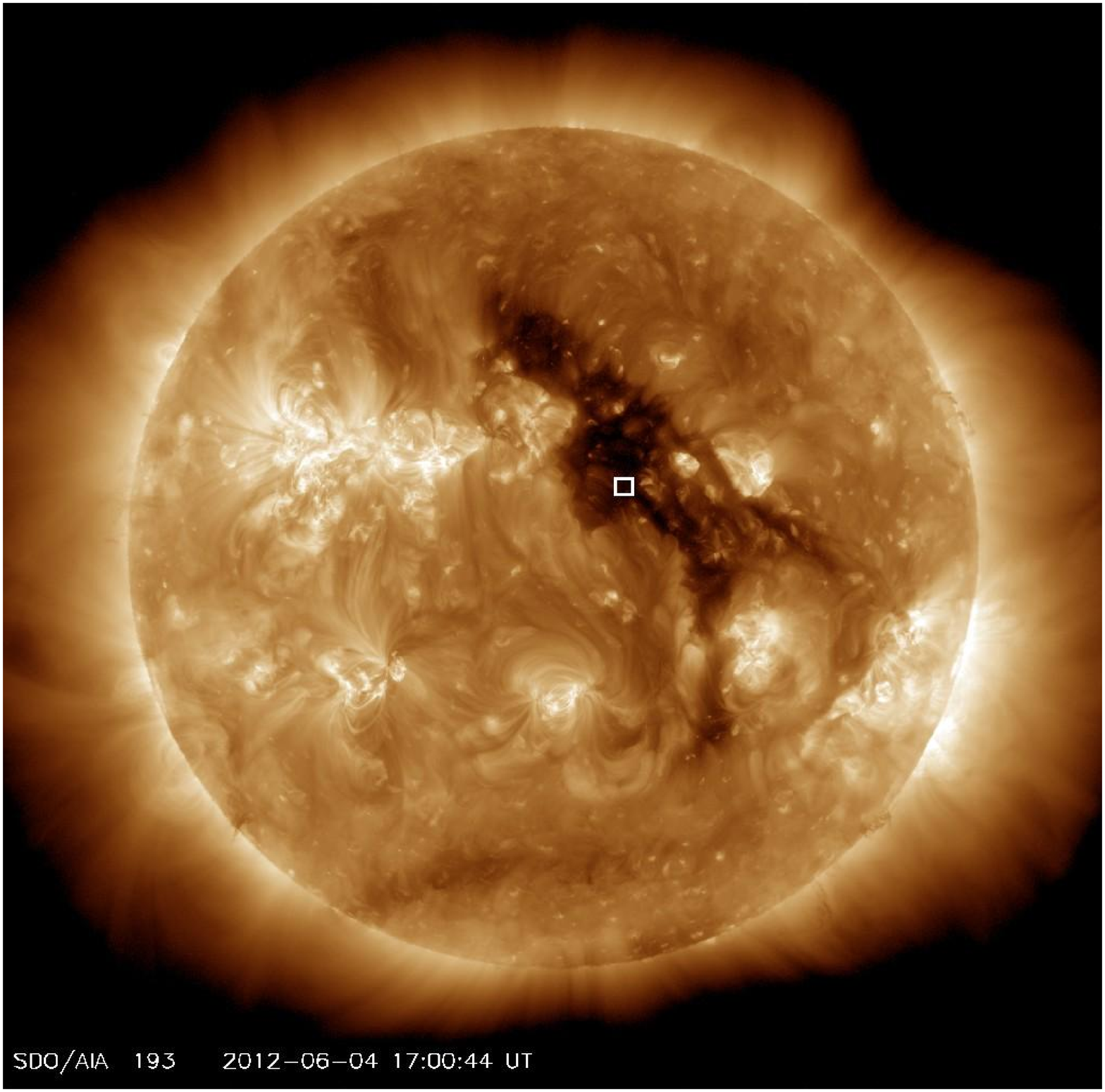}}
\caption{SDO/AIA/193\AA\ images showing the location of the two CHs observed on
2011 August 12 (left) and 2012 June 4 (right). The white boxes indicate the
location of the NST FOV. }
\label{fig1}
\end{figure}
%#####################################################################

\begin{figure}[!ht]
\centerline{
\epsfxsize=6.0truein \epsffile{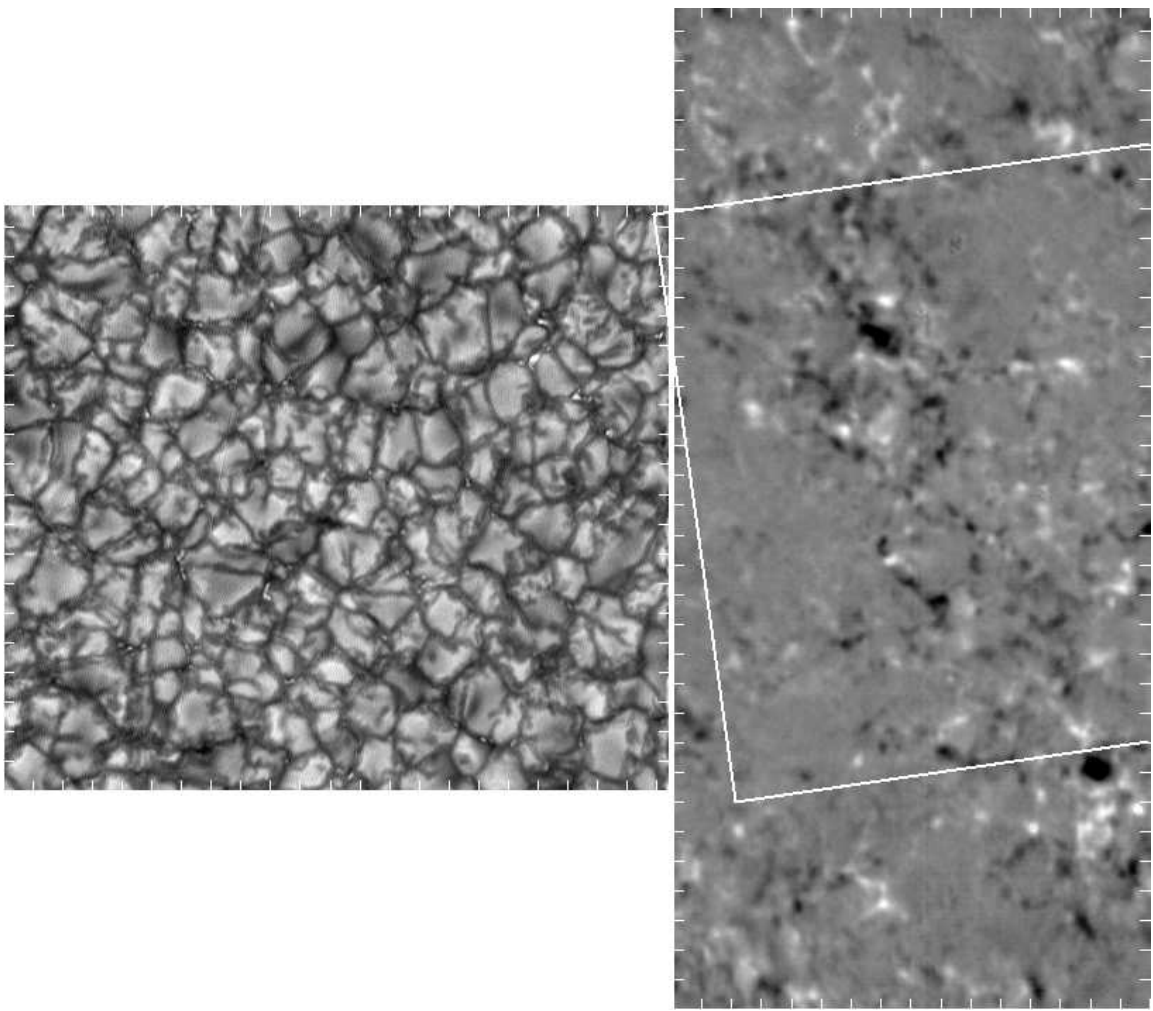}}
\caption{Left - one of the 659 images of CH 2012-06-04 obtained with the NST/TiO
filter. Right - the co-temporal NST/IRIM LOS magnetogram. 
The white lines outline three boundaries of the FOV/TiO.
 The magnetogram is scaled from -100 G (black) to 100 G (white).
The magnetogram and the granulation are shown in the same scale. The ticks 
separate a distance of 1~Mm.}
\label{fig2}
\end{figure}

The solar granulation data were acquired with a broad-band TiO filter centered
at 705.7 nm with a band-pass of 1 nm \citep[for details see][]{Abramenko+2012}.
The observations were aided with an adaptive optics (AO) system
\citep{Denker+AO-2007,Cao+2010-IRIM+AO}. To take maximal advantage of the AO
system, only the central part of the entire FOV (77$''\times$77$''$) was
utilized. The pixel scale of the PCO.2000 camera, 0.${''}$0375, was 2.9 times
smaller than the telescope diffraction limit of 77 km. The Kiepenheuer-Institut
f{\"u}r Sonnenphysik's software package for speckle interferometry
\citep[KISIP,][]{Woger-Luhe-kisip-2007} was applied to the 70 best images
selected from a burst of 100 images to produce one speckle reconstructed image
at the diffraction limited resolution. After destretching and sub-sonic
filtering \citep{Abramenko+2011-Diff} of speckle reconstructed images, the final
CH 2011-08-12 (CH 2012-06-04) data set consisted of 82 (659) images. The final
time cadence was 12 and 13~s, respectively. One-two minutes gaps are sometimes
present in the data. An example of a TiO/NST image of granulation is shown in
Figure \ref{fig2} ~(left).

Spatial fluctuations of the {\it transverse} component of the magnetic field
inside a CH are of primary interest in this paper. Unfortunately, these kind of
data are less plentiful, and the accuracy of measurements is much lower than
that for the line-of-sight (LOS) data, especially in the quiet sun and in
coronal holes. With this in mind, we address the problem in the following way.
We utilized Hinode/SOT/SP data acquired inside a quiet sun region at the solar
disk center on 2007, March 10. The processed data of the magnitude of the
transverse magnetic field $B_{app}^T$ and the corresponding LOS magnetic
field $B_{app}^L$ were kindly provided to us by Dr. B. Lites. The SOT/SP slit
scanned an area of 304$''\times$164$''$, or 220$\times$120 Mm. The pixel size
was 0.$''$150$\times$0.$''$160. The data acquisition, processing and noise
issues are described in detail by \citet{lites2008}. The maps of the magnetic
field components are shown in their Figure 2. We consider the $B_{app}^T$ data
set as the primary source for deriving the characteristic length of the
transverse magnetic fluctuations. For comparison, we also calculated the
$\lambda$, $\varsigma$, and $L$ from the LOS component $B_{app}^L$. 

We also utilized LOS magnetograms obtained for CH 2012-06-04 with the
near-infrared imaging magnetograph \citep[IRIM,][]{Cao+2010-IRIM+AO} installed
on the NST. The spectro-polarimeter uses the Fe~I spectral line at 1.56~$\mu$m
and is based on a 2.5~nm interference filter, a 0.25~\AA~ birefringent Lyot
filter
\citep{Lyot_Filter}, and a Fabry-P\'{e}rot etalon and provides a
bandpass of $\sim$0.01~nm
over a FOV of $50'' \times 25''$ with a pixel scale of 0.$''$098. Based on the
advantages of the near IR window and the NST adaptive optics, IRIM provides
solar spectro-polarimetry data with Zeeman sensitivity of 10$^{-3} I_c$,
diffraction limited resolution of 0$''$.2, and time cadence of 45~s (for full
Stokes profiles). The raw IRIM data were processed to take into account dark and
flat field corrections and polarization calibration 
\citep{Cao+2010-IRIM+AO,Goode-Cao+2011,Cao+2011}. The linear polarization signal
(the source of the transverse magnetic field data) from IRIM is very weak in
CHs, therefore noise issues precluded scientific usage of it, while the LOS
magnetic field component was reliably derived by using the weak-field
approximation. The best 34 LOS magnetograms (Figure \ref{fig2}, right) were
utilized here to estimate the characteristic length of $b_z$.
%#####################################################################
%#####################################################################

\section{Transverse Velocities Acqusition}

To derive transverse velocities in the photosphere, we applied the Local 
Correlation Tracking (LCT) routine to the time series of solar granulation.
The routine was first proposed by \citet{November-1988-LCT} and later modified  
and improved
\citep[e.g.,][]{Title-1995-LCT,Hurlburt-1995-LCT,strous1996,Welsch+2007-LCT, 
Chae-Sakurai-2008-LCT,Matloch+2010-LCT,Verma-Denker-2011-LCT,Verma+2013-Arxiv}.

\citet{Matloch+2010-LCT} tested the performance of the LCT on simulated data 
for quiet-sun granulation. The used LCT code is described in \citet{Welsch+2007-LCT}. 
The authors found that the LCT-derived velocities were 
approximately two times smaller than those derived directly from the 
simulations. The agreement between the simulated and LCT velocities was found 
to depend on the depth in the simulation box. The correlation reaches maximum 
($\sim$0.65) at the depth close to the optical depth of $\tau =1$, i.e., near 
the continuum formation level.

\citet{Verma-Denker-2011-LCT} and
\citet{Verma+2013-Arxiv}
presented a 
comprehensive study of accuracy and reliability of the LCT technique by 
applying it to {\it Hinode} G-band images. They demonstrated 
\citep{Verma-Denker-2011-LCT} that the choice of three essential input
parameters (time 
cadence, $\delta t$, accumulation interval, $T$, and the FWHM of a Gaussian 
sampling window, $W$) may significantly affect the result. 
\citet{Verma+2013-Arxiv} evaluated the LCT technique using
CO5BOLD simulations of
solar 
granulation \citep{Freytag+2012-CO5BOLD}. Note, that the spatial and temporal 
resolution (28~km pixel scale and 10~s lowest time cadence) of the simulated 
data set are comparable to those of the observed NST set analyzed here (27 km 
pixel scale and 12-13 s time cadence). 
\citet{Verma+2013-Arxiv}
found that, when 
multiplied by a factor of three, the frequency distribution of the LCT 
velocities demonstrates a very good agreement with the distribution of the 
velocities derived directly from the simulated set 
\citep[see Figure 13
in][]{Verma+2013-Arxiv}.
Furthermore, Figure 7 in \citet{Verma+2013-Arxiv} 
suggests that the time cadence of 10-30~s and the apodization window of 
600-1000~km are the best choice for analysis of the inter-granular dynamics. 
Interestingly, the narrower window of 400~km results in diminishing of the LCT 
velocities and degradation of the fine structures in the velocity pattern, so 
that the LCT technique fails to capture the inter-granular dynamics there.

Figures 4 - 6 in \citet{Verma+2013-Arxiv}, as well as
Figure 3 in 
\citet{Verma-Denker-2011-LCT} demonstrate that the extension of the 
accumulation interval $T$ results in diminishing of the LCT velocities as 
well. The authors found that in order to reveal the persistent (e.g., meso-
granular and/or super-granular) flows in the solar granulation, the 
accumulation time should be no less than approximately 20 minutes. One might 
conclude that shorter time intervals should be employed to reveal inter-
granular dynamics.

Here we focus on analysis of inter-granular dynamics, i.e., on proper motions 
of granule edges, mini-granules \citep{Abramenko+2012}, inter-granular bright 
points (BPs) and filigree-like features. According to MHD simulations 
\citep{Stein-2013-private-c} and NST observations, these features visible 
inside inter-granular lanes are mostly associated with strong quiet Sun 
magnetic fields. It is thought that Alfven waves are generated by rapid 
displacements of magnetic footpoints driven by these inter-granular flows and 
the energy-containing scale should be related to this kind of motions. We will 
choose the LCT configuration parameters based on the above considerations and 
the results of \citet{Verma-Denker-2011-LCT} and \citet{Verma+2013-Arxiv}.

The version of the LCT used here was elaborated by \cite{Strous-1994-thesis} 
and \cite{strous1996} and it is utilizing ANA software package \footnote{see 
more at \url{http://ana.lmsal.com/ana/}}. We start with choosing  the best 
time cadence $\delta t$ using the CH-2012-06-04 data recorded with time 
cadence of 13~s. We generated two additional sets with $\delta t \approx$ 39 
and 65~s by selecting each third and fifth image, respectively. In this case, 
with $T=$10~min and $W=$540~km, we obtained the following values of the square 
root of the averaged squared velocities:
1.21~km s$^{-1}$ for  $\delta t$=13~s,
0.61~km s$^{-1}$ for  $\delta t$=39~s, and
0.26~km s$^{-1}$ for  $\delta t$=65~s.
We thus concluded that the first estimate of the LCT velocity (1.21~km s$^{-1}$ 
derived for $\delta t$=13~s) is the best choice because it is the largest one 
out of the three and the closest one to the estimates derived by \citet[][see 
Figure 8]{Verma+2013-Arxiv} for the range of $\delta t$ =
10-30~s and $W$ = 
600 - 1000~km.
 
The next critical parameter is the FWHM of the Gaussian sampling window $W$. 
According to \citet{Verma-Denker-2011-LCT} and \citet{Verma+2013-Arxiv}, the 
choice of $W$ must be guided by the spatial scales of the studied dynamic 
events. In case of the inter-granular scale events, the window should not 
exceed the typical size of a granule, i.e, $\sim$1000~km. We thus tested three 
choices of $W$: 10$\times$10 pixels (270~km in linear extent), 20$\times$20 
pixels (540~km), and 40$\times$40 pixels (1090~km) for which we obtained the 
following square root of the averaged squared velocities: 1.02~km s$^{-1}$ for  
$W=$ 270~km, 1.21~km s$^{-1}$ for $W=$ 540~km, and 1.11~km s$^{-1}$ for $W=$ 
1090~km. Again, we chose the $W=$ 540~km value since it ensures the largest 
detected velocities.

Finally, the accumulation (averaging) time interval $T$ should be shorter 
than 20 minutes. Similarly, four choices of $T$ were tested: 2, 5, 10, and 
20~min. The following square root of the averaged squared velocities were 
obtained: 1.71, 1.40, 1.21, and 1.06~km s$^{-1}$, respectively. Qualitatively, 
the tendency here agrees with that reported by \citet{Verma-Denker-2011-LCT} 
and \citet{Verma+2013-Arxiv}: the averaged velocity decreases as $T$ increases. 
However, there is no indication which one might be closer to the real value. 
The possible answer is that all of them are true, and they simply reflect the 
dynamics on different time scales. We, therefore, decided to proceed using all 
of the above accumulation intervals.

As we saw above, underestimation of the LCT velocities by 2 to 3 times is an 
unavoidable shortcoming of the LCT technique. However, in this particular 
study, we are focused on calculation of the correlation functions 
characterizing LCT velocities structures rather than on the magnitude of the 
velocities. According to the general expression of the correlation function 
(see Eq.\ref{Br} below), multiplying the velocity (or squared velocities) by a 
constant factor does not change the correlation function (the factor evenly 
appears in the numerator and the denominator of the right-hand part of Eq.
\ref{Br}). 

Note, that the TiO spectral line is formed in the quiet sun at optical depths 
of $\tau=1$, i.e., near the continuum level 
\citep{Abramenko+2012,Berdyugina_2003}, where the LCT is expected to perform 
in the most reliable way \citep{Matloch+2010-LCT}.

\section{ Different Methods to Derive the Characteristic Length}

Our main goal is to find, following \citet{Batchelor-book}, $''$convenient
measures of the linear extent of the region within which velocities
[as well as squared velocities and magnetic field] are appreciably correlated$''$. 
For an arbitrary two-dimensional scalar field $v({\bf r})$, a general definition
of the correlation function $B({\bf r})$ reads as
\citep{Jenkins_Watts,Monin_Yaglom}
\begin{equation}
B(r) = \langle (v({\bf x} + {\bf r})-\langle v\rangle) 
\cdot (v({\bf x})-\langle v\rangle)\rangle /\sigma^2,
\label{Br}
\end{equation}
where ${\bf r}$ is a separation vector,  ${\bf x} \equiv (x,y)$ is
the current point on an image, and $\sigma^2$ is the variance of $v$. Angle
brackets denote averaging over the area of an image.
Based on $B({\bf r})$, the characteristic length of the field $v$ can be
defined as a distance where some correlation in $v$ holds. 

We will use three different approaches for deriving the characteristic length from
the correlation function $B(r)$. For a noise-free smooth image, measured on a
large enough area, all of our approaches produce the same magnitude for the
characteristic length. However, in case of real data, the estimates from different
methods might differ. We will consider the results obtained from these
approaches as independent estimates of the characteristic length for a given
image.

The first approach is to determine the integral scale \citep{Batchelor-book}
\begin{equation}
\lambda=\int_0^{r_{max}} B(r) dr,
\label{integ}
\end{equation}
which has also been used in the formulation/derivation of the turbulence
transport model of \citet{Zank+2012}. Here the integration is supposed to be
performed over all scales up to infinity. However, in the case of real data, at
scales exceeding a certain large distance, data noise might cause a
non-diminishing $B({\bf r})$, thus resulting in an artificially large integral
scale. To mitigate this problem, the signal in pixels was assigned to be zero
when the absolute value of the measured signal was below  one standard
deviation. Moreover, we noticed that in our data sets an increase in correlation
is observed at scales larger than approximately 5 Mm, which is caused by the
presence of adjacent features in a studied field. To avoid an overestimation of
the integral length, we adopted $r_{max}$ in Eq. \ref{integ} to be 5~Mm.

The second method is to approximate the correlation function near the origin 
\citep{Hinze,Monin_Yaglom,feder1989}:
\begin{equation}
B(r)= Const \cdot exp(-r/\varsigma).
\label{dzeta}
\end{equation}
Thus, when  $B(r)$ is fitted with an exponential function, the correlation
function drops by $e$ times at a scale $\varsigma$. This scale is called the
correlation length in the percolation \citep{feder1989} and in second-order
phase transition \footnote{
http://staff.science.nus.edu.sg/~parwani/c1/node52.html} theories.

Finally, the characteristic length can also be determined via the $e$-folding
scale of $B(r)$ without any approximation of the latter (the scale, where the
{\it measured} correlation function drops by $e$ times).  We denote this measure
as $L$.

The parameters introduced above $\lambda$, $\varsigma$, and $L$ are considered
as proxies for the characteristic length, and were calculated for all data sets.

\section{Results} 

\subsection{Characteristic Length of Transverse Velocity Fluctuations}

\begin{figure}[!ht]
\centerline{
\epsfxsize=5.0truein \epsffile{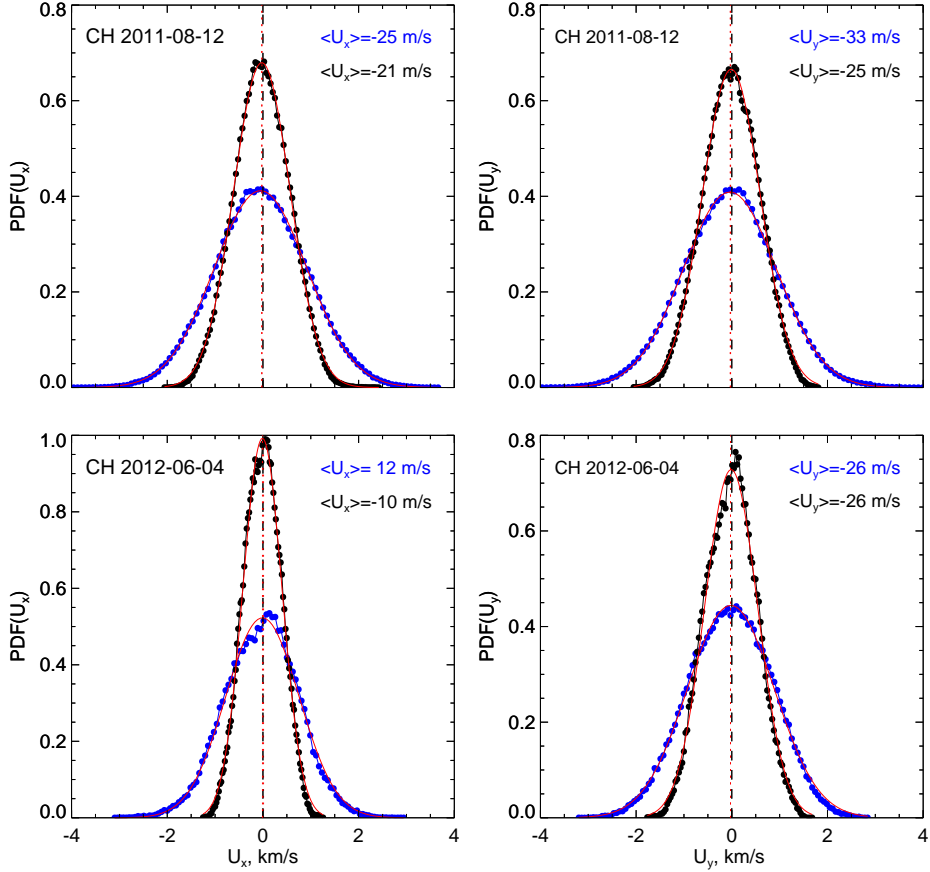}}
\caption{Probability distribution functions (PDFs) of the transverse velocities,
$U_x$ (left) and $U_y$ (right) derived from a velocity map for CH 2011-08-12
(top, 734 x 734 data points) and CH 2012-06-04 (bottom, 351 x 351 data points).
PDFs, obtained with the accumulation time interval $T=$2 (20) min are shown with
blue (black) circles. The red curves are the best Gaussian fit to the data
points. The mode of the Gaussian fit is shown for each data set with the same
color code. The Gaussian mode deviates from zero by less than 35 m
s$^{-1}$ for all flow maps. }
\label{fig3}
\end{figure}

\begin{figure}[!ht]
\centerline{
\epsfxsize=3.0truein \epsffile{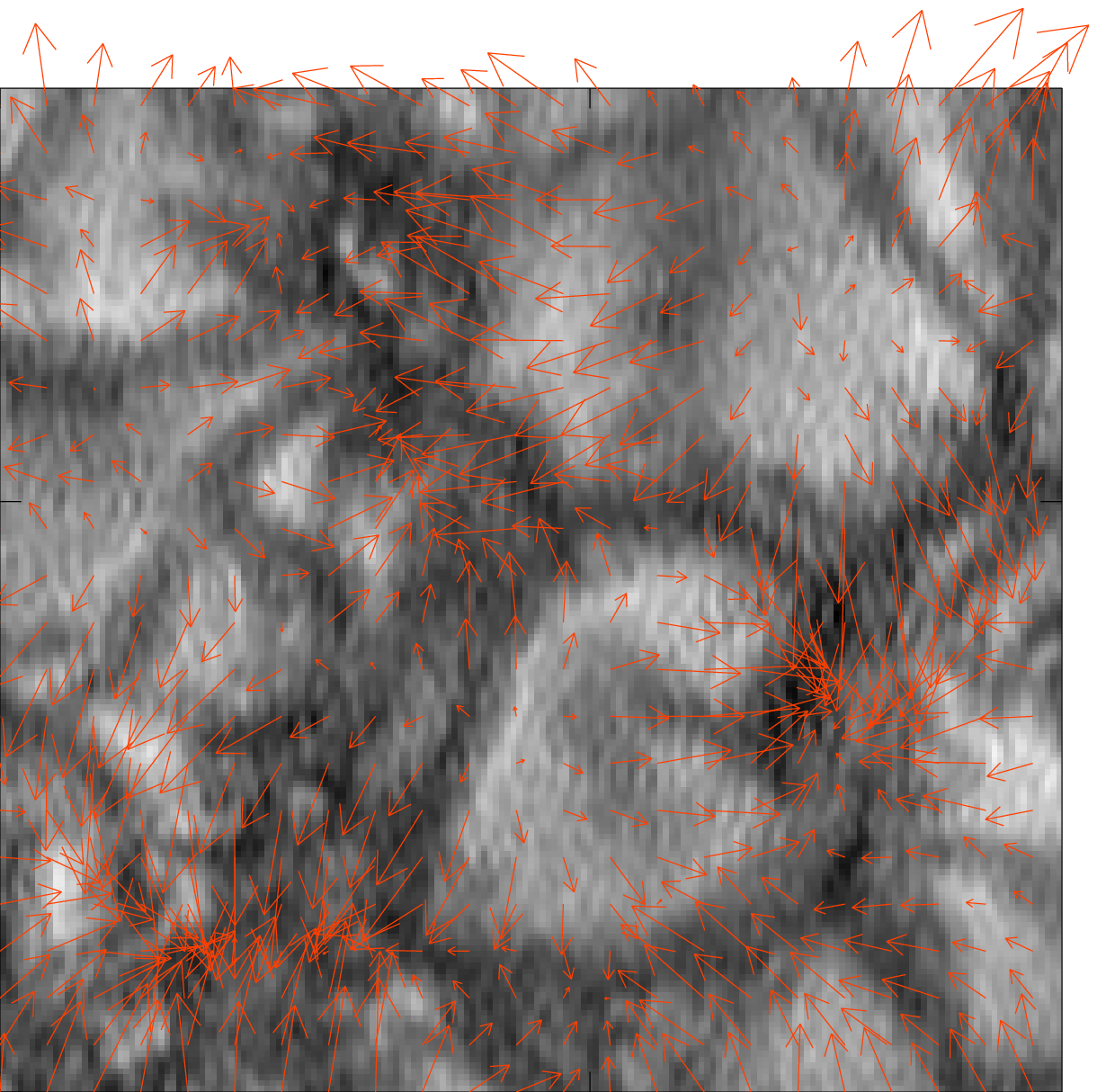}
\epsfxsize=3.0truein \epsffile{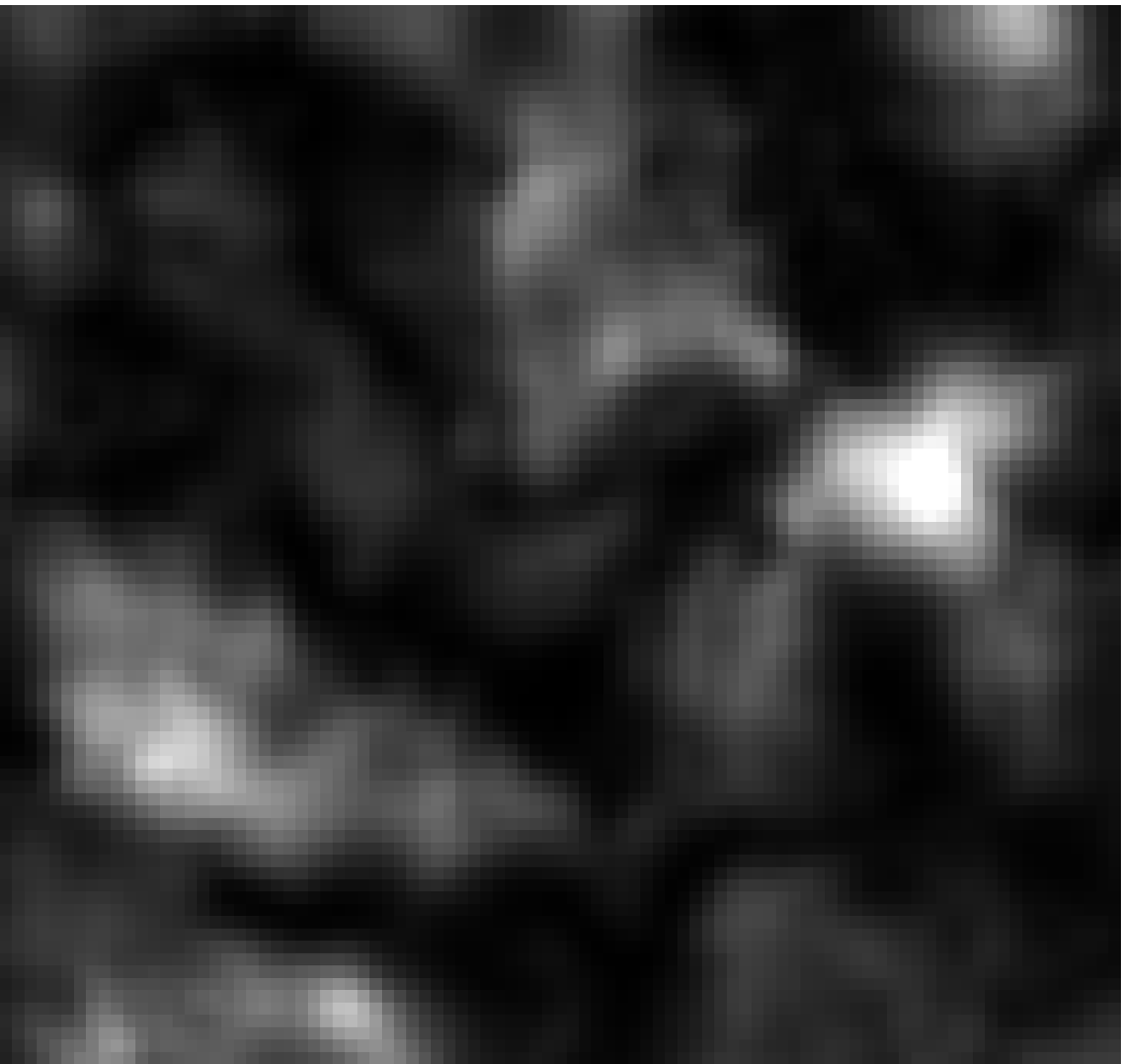}}
\caption{{\it Left} - NST/TiO image of solar granulation inside CH 2012-06-04
overplotted with the corresponding velocity fluctuations map (arrows) obtained
with $T=$10 min. The image size is 4.9$\times$4.6~Mm. The
length of the arrows is in proportion to the velocity, and the longest arrows
correspond to 2.2 km s$^{-1}$. {\it Right} - the corresponding map of squared
transverse velocity fluctuations, $u^2$, saturated at 4 km$^2$s$^{-2}$.}
\label{fig4}
\end{figure}

Dividing the resulting LCT displacements in each pixel by the accumulation time
$T$, we obtained components of the transverse velocities $U_x$ and  $U_y$ (flow
maps). We  then computed the probability distribution functions (PDFs, Figure
\ref{fig3}) of $U_x$ and $U_y$ for each flow map, 
%#####################################################################
%#####################################################################
which showed that alignment of the images in the data sets was very good, and
there are no large-scale systematic shifts, so that residual large-scale
velocities (mean velocities) are negligible. Indeed, the PDFs of the $U_x$ and
$U_y$ components are well centered at zero and are distributed in accordance
with a Gaussian function. The mean velocity components  $\langle U_x \rangle$
and  $\langle U_y \rangle$ do not exceed 35 m s$^{-1}$ in any flow map. After
subtraction of the mean velocities from the flow maps, we obtained the velocity
fluctuations maps $u_x$ and $ u_y$. A fragment of a $u_x, u_y$-map is shown in
Figure \ref{fig4}, {\it left}. Centers of large granules are usually places
of slow transverse motions, whereas the majority of long arrows are located near
the periphery of large granules and inside inter-granular lanes, thus supporting
the idea of enhanced dynamics in inter-granular lanes and proving the capability
of the LCT to detect it.
%#####################################################################
%#####################################################################

We first focus on the statistical properties of the squared amplitude $u^2 =
(u_x^2 + u_y^2)$ of the $(u_x, u_y)$-vector, which is a positive scalar that
characterizes the kinetic energy of photospheric random motions driven by
convection and turbulence. Therefore, the $u^2$ -maps represent an entity in
which we are interested (see Eqs. 2-6 in \citet{Zank+2012}). In the right panel
of Figure \ref{fig4}, we show a $u^2$-map calculated from the flow map shown in
the left panel. Patches of enhanced kinetic energy (white areas) are
intermittent with dark voids. Our aim is to estimate the characteristic length
of these patches.    

\begin{figure}[!ht]
\centerline{
\epsfxsize=6.0truein \epsffile{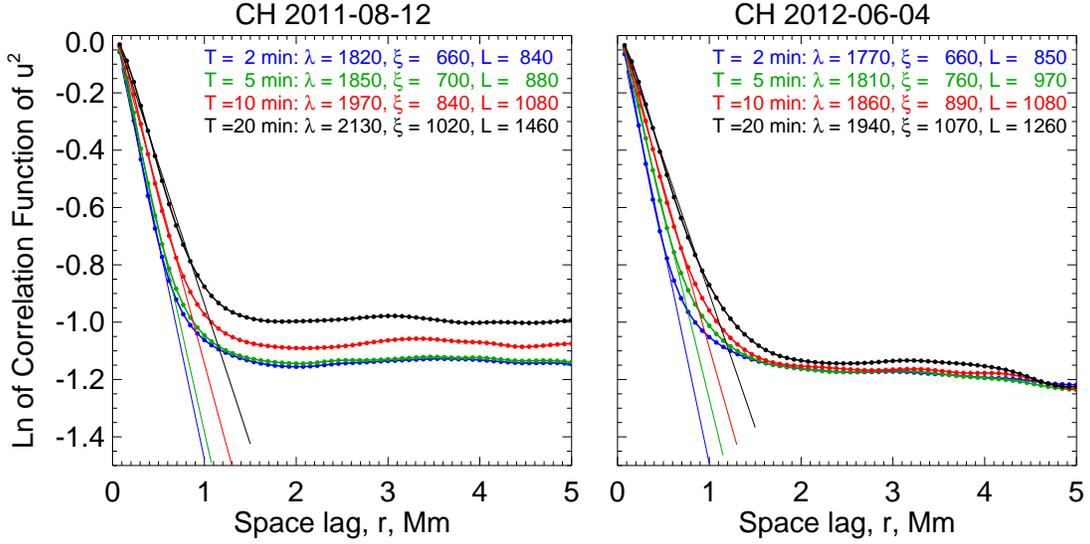}}
\caption{Plotted is the natural logarithm of the correlation function of $u^2$
versus the spatial lag $r$ for the two CHs. The circles denote the data points
and the straight lines are best linear fits to the first 8-10 data points used
to derive $\varsigma$ according to Eq. \ref{dzeta}. The color code refers to the
accumulation time intervals of 2 (blue), 5 (green), 10 (red), and 20 (black)
minutes. The values of the integral length $\lambda$ the correlation length
$\varsigma$ and the $e$-folding length $L$ in km are shown.}
\label{fig5}
\end{figure}

Figure \ref{fig5} shows the correlation functions of $u^2$ structures determined
with different accumulation intervals $T$. All the proxies of the
characteristic length $\lambda, \varsigma$, and $L$ (see Sec. 4) show an
increase with larger $T$, which seems to be expected because various small-scale
velocity patterns tend to be smeared as they are averaged over time longer time
intervals (Table 1.) The estimate of the Batchelor integral length $\lambda$
is the largest. For all cases, $\varsigma$ is less than $L$. Based on the three
proxies, the characteristic length of $u^2$ varies in the range of 660 - 2130~km
with an average value of 1260$\pm$500~km.

\begin{table}[!ht]
\caption{\sf Characteristic lengths of the squared transverse velocity fluctuations, $u^2$}
\footnotesize
\begin{center}
\begin{tabular}{lrrr}
\hline
    &  $\lambda$, km &  $\varsigma$, km &$L$, km \\      
        
\\

\hline
CH 2011-08-12 &  &  &  \\
\hline
\hline
 $T$= 2 min& 1820 & 660 &  840 \\
\hline
 $T$= 5 min& 1850 & 700 &  880 \\
\hline
 $T$=10 min& 1970 & 840 &  1080 \\
\hline
 $T$=20 min& 2130 & 1020 &  1460 \\
\hline
\hline
CH 2012-06-04 &  &  &  \\
\hline
$T$= 2 min& 1770 & 660 &  850 \\
\hline
$T$= 5 min& 1810 & 760 &  970 \\
\hline
$T$= 10 min& 1860 & 890 &  1080 \\
\hline
$T$= 20 min& 1940 & 1070 &  1260 \\
\hline
\hline
$\langle \lambda_{\bot} \rangle $ & 1890$\pm$120 & 830$\pm$160 & 1050$\pm$220 \\

\hline
\end{tabular}
\end{center}

\end{table}

%#####################################################################
%#####################################################################

%#####################################################################
\begin{figure}[!ht]
\centerline{
\epsfxsize=6.0truein \epsffile{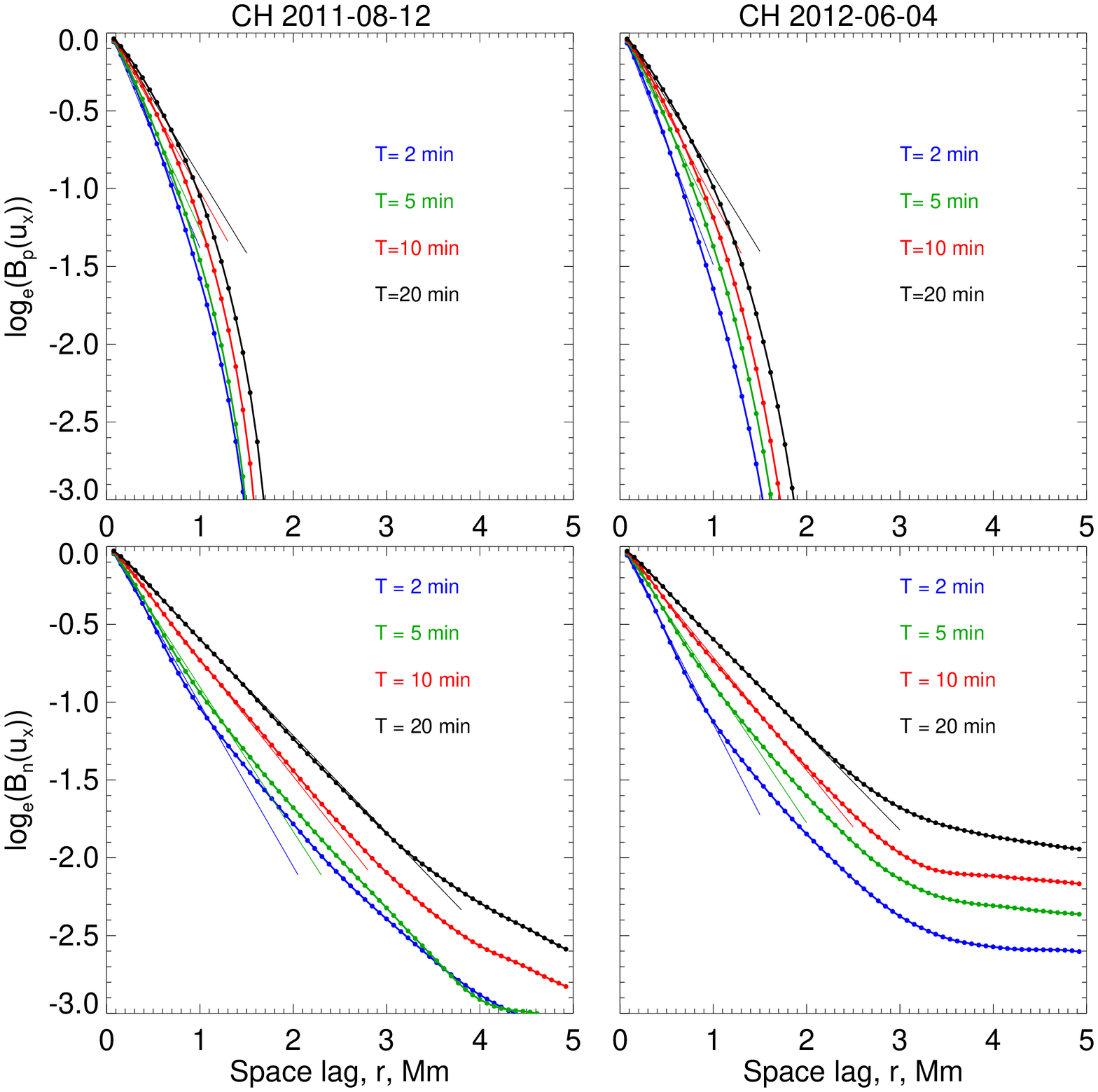}}
\caption{Plotted is the natural logarithm of the parallel (upper row) and normal
(bottom row) correlation functions for $u_x$ versus the spatial lag $r$. The
notation is the same as in Figure \ref{fig5}. }
\label{fig6}
\end{figure}
%#####################################################################

We then calculated correlation functions of the transverse velocity components
$u_x$ and $u_y$. For a 2D vector field (unlike a scalar field), one has to
compute two correlation functions \citep{Batchelor-book}: one of them, $B_p$, 
is parallel to the separation vector $\bf {r}$, and the other one, $B_n$, is
normal to $\bf {r}$. To simplify the calculations without sacrificing the
quality, we computed $B_p(u_x)$ assuming that $r$ varies along the $x$-axis
only. $B_n(u_x)$ was computed with $r$ varying along the $y$-axis. The
correlation functions for $u_x$ are shown in Figure \ref{fig6}. Similarly, we
computed both correlation functions for $u_y$, which were very similar to that
derived for $u_x$, so their plots are not shown.
Obtained from the plots averaged (over $T$  and both
CHs) values of $\lambda,\varsigma$, and $L$ are gathered in Table 2.  

\begin{table}[!ht]
\caption{\sf Averaged characteristic lengths of the components $u_x$ and $u_y$
from the parallel $B_p(r)$ and normal  $B_n(r)$
correlation functions }
\footnotesize
\begin{center}
\begin{tabular}{lrrrrrr}
\hline
    &  $\lambda$, km &  $\varsigma$, km &$L$, km   $\mid$&  $\lambda$, km & $\varsigma$, km &$L$, km\\ 
\hline     
    &                 &       $u_x$     &          $\mid$&                &    $u_y$        &       \\    
\hline
$\langle \lambda_{\bot} \rangle $ for $B_p(r)$ & 650$\pm$100 & 840$\pm$140 & 830$\pm$120 $\mid$& 800$\pm$190 & 860$\pm$160& 850$\pm$140\\
\hline
$\langle \lambda_{\bot} \rangle $ for $B_n(r)$& 1340$\pm$230 & 1240$\pm$290 & 1260$\pm$300  $\mid$& 1430$\pm$250 & 1270$\pm$320& 1300$\pm$320\\
\hline

\end{tabular}
\end{center}

\end{table}

The averaged integral length $\lambda$ derived for both CHs using $u_x$
and $u_y$ from the parallel correlation functions $B_p(u_x), B_p(u_y)$ is
720$\pm$160~km, whereas $\lambda$ calculated from the normal correlation
functions is 1380$\pm$230~km. Both are smaller than $\lambda$ obtained from
$B(u^2)$. This can be understood from the non-negative nature of the $u^2$ field
which gives an extended positive tail for $B(u^2)$. This is not the case for the
parallel and normal correlation function of the alternating-sign fields of $u_x$
and $u_y$. The correlation length $\varsigma$ and the $e$-folding length $L$
from $u_x, u_y$ agree within the error bars with that derived from $u^2$. 

%#####################################################################
\begin{figure}[!ht]
\centerline{
\epsfxsize=6.0truein \epsffile{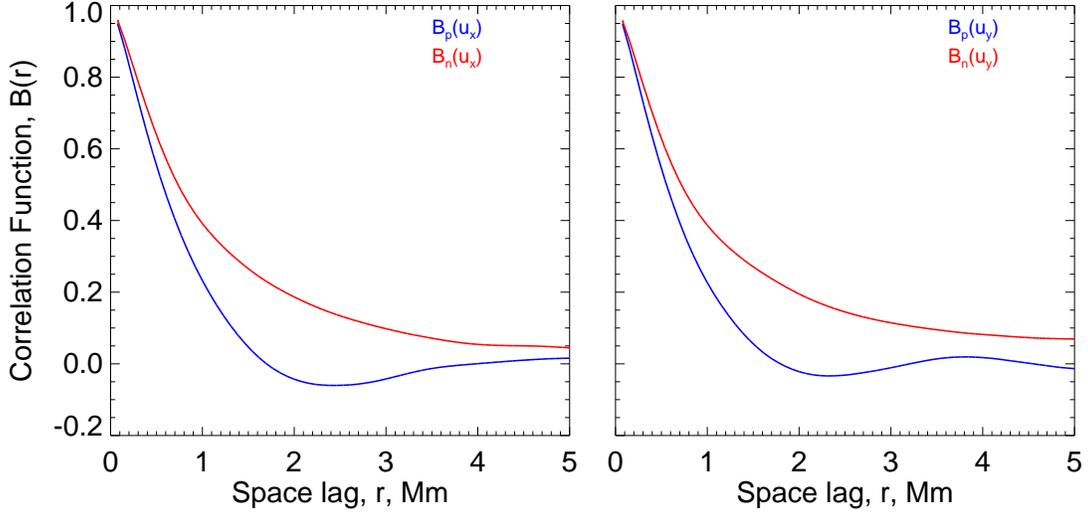}}
\caption{An example of the parallel (blue) and normal (red) correlation
functions calculated from $u_x$ (left) and $u_y$ (right) components of the
transverse velocity fluctuations. The data for CH-2011-08-12 obtained with the
accumulation time of $T=$ 5 min are used for these plots.
}
\label{fig7}
\end{figure}
%#####################################################################

We find that the behaviour of the parallel and normal correlation functions is
different (see Figure \ref{fig7}). Whereas the normal correlation function $B_n$
is positive for all scales, the parallel function changes sign, always situated
below the normal correlation function; i.e., $B_p(r) < B_n(r)$. This situation
holds for all the analysed data sets and for both velocity components $u_x$ and
$u_y$. On average, the parallel correlation function has a characteristic length
of 800$\pm$160~km, whereas the characteristic length derived from the normal
correlation function is 1310$\pm$280~km. Note that in the case of homogeneous
isotropic hydro-dynamical turbulence, the mutual behavior of the parallel and
normal correlation functions of velocity fluctuations is expected to be quite
the opposite, i.e., the parallel correlation function exceeds the normal
correlation function at all scales: $B_p(r) > B_n(r)$
\citep{Batchelor-book,Monin_Yaglom}. Obviously, the condition of homogeneous
isotropic turbulence is not met in the  solar photosphere. Moreover, the
presence of magnetic fields could also contribute in the observed peculiar
relationship between the correlation functions. 

\subsection{Characteristic Length of the Magnetic Field Fluctuations}

%#####################################################################
\begin{figure}[!ht]
\centerline{
\epsfxsize=6.0truein \epsffile{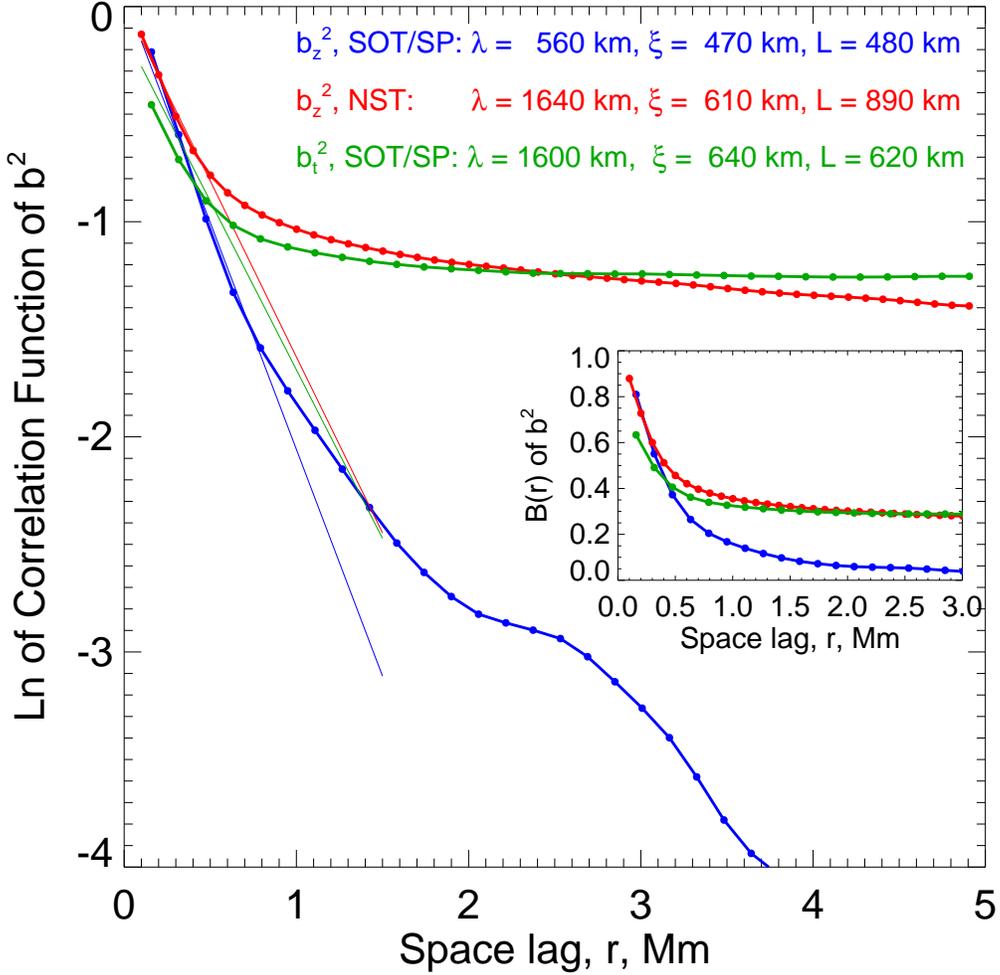}}
\caption{Plot of the natural logarithm of the correlation function of $b^2$
versus the spatial lag $r$. Circles denote the data points, and the straight
lines are the best linear fits to the first data points used to derive the
values of $\varsigma$ from Eq. \ref{dzeta}. The insert shows this plot with
linear axes. }
\label{fig8}
\end{figure}
%#####################################################################
The correlation functions from the squared magnetic field components are shown
in Figure \ref{fig8}. The most interesting to us is the correlation function of
the squared transverse magnetic field component $b_t^2$ plotted with the green
line. On scales below approximately 0.5~Mm, the three correlation functions are
similar. On larger scales, the NST $b_z^2$ function (red line) agrees well with
that derived from $b_t^2$. (The estimates for $\lambda, \varsigma$, and $L$
derived from the green and red curves are similar too.) It is plausible that the
very low noise level in the $B_{app}^L$ data \citep[about 2-3
Gauss,][]{lites2008} is the reason why on scales larger than 1~Mm the blue curve
in Figure \ref{fig8} is much lower than the green and red curves. Thus, the
correlation functions from the squared LOS component of the magnetic field can
be used as an appropriate proxy to estimate the the correlation function of the
squared transverse component. The characteristic length of the squared
transverse magnetic field fluctuations varies in a range of 617 - 1600~km with
an average value of 950$\pm$560~km.

%#####################################################################
\begin{figure}[!ht]
\centerline{
\epsfxsize=6.0truein
\epsffile{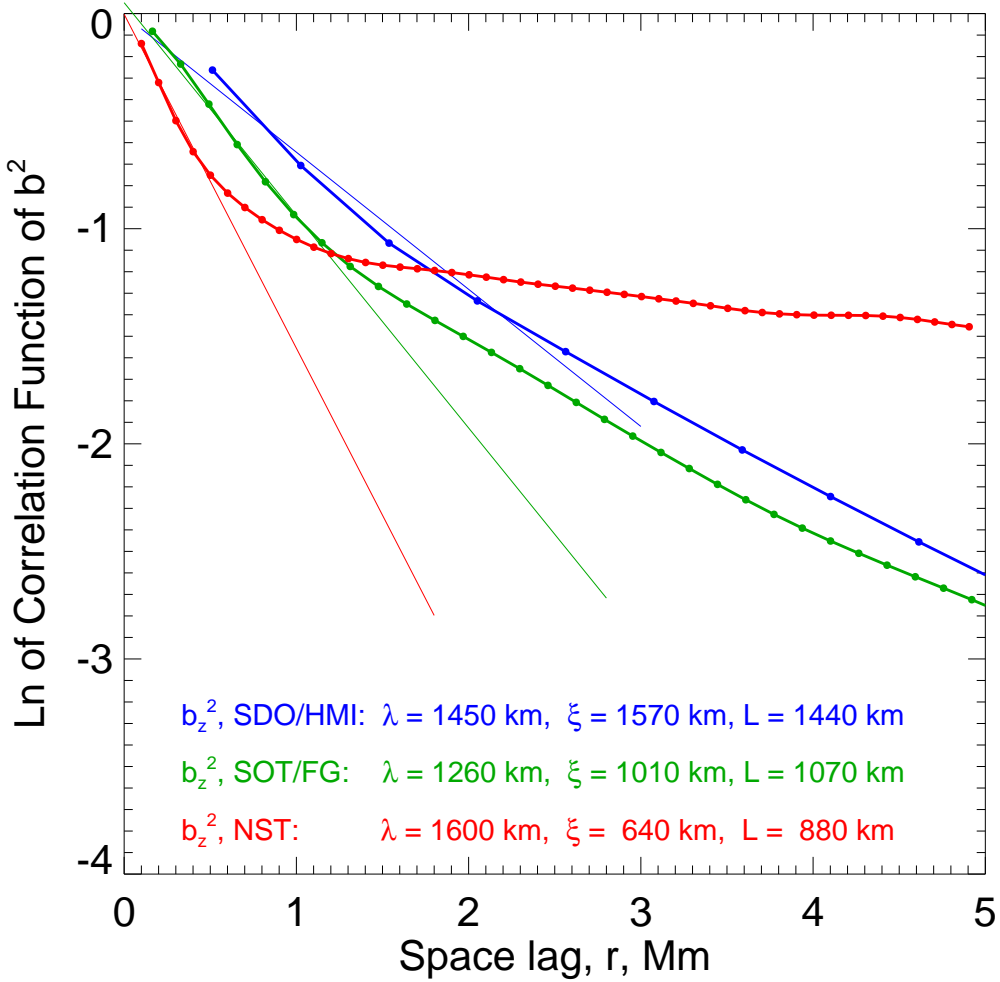}}
\caption{Plot of the natural logarithm of the correlation function of $b_z^2$
versus the spatial lag $r$ for the same area on the Sun (the CH 2011-08-12)
but derived from data acquired with different instruments. The notation is the
same as in Figure \ref{fig8}. }
\label{fig9}
\end{figure}
%#####################################################################

When comparing the characteristic lengths derived for $u^2$ with those derived
for $b_t^2$, we find that for the integral length
$\lambda(u^2)/\lambda(b_t^2)\approx$1.2 for the correlation length
$\varsigma(u^2)/\varsigma(b_t^2)\approx$1.3 and for the $e$-folding length
$L(u^2)/L(b_t^2)\approx$ 1.7. Thus, the characteristic scale of the squared
velocity fluctuations is larger than that of the squared transverse magnetic
field.  

In Figure \ref{fig9}, we compare the $b_z^2$-correlation functions derived from
magnetic field data obtained with SDO/HMI (pixel scale 0.$''$5), {\it Hinode}
SOT/FG (pixel scale 0.$''$16), and NST/IRIM (pixel scale 0.$''$098) instruments
for the same area inside the CH-2011-08-12. The figure demonstrates that the
correlation function becomes narrower and produces smaller estimates of the
characteristic length as the spatial resolution improves.  

\section{Conclusions and Discussion}

Using high spatial (0.$''$1) and temporal (12-13~s) resolution time series of
solar granulation acquired inside two CHs with the NST at BBSO, we computed the
characteristic (correlation) length of the transverse velocity fluctuations
$u^2$.  The corresponding length of the transverse magnetic field fluctuations
$b_t^2$ was derived from a unique and high quality Hinode SOT/SP data set which
covered a large area of 220$\times$ 120 Mm. 

The characteristic length of the energy containing structures $u^2$ and
$b_t^2$ was derived from correlation functions of  $u^2$ and $b_t^2$ by using
three independent methods. These were calculated from the Batchelor integral
scale $\lambda$, the correlation length $\varsigma$, and the $e$-folding
length $L$.

For $u^2$-structures, Batchelor integral scale $\lambda$ varies in the range
of 1800 - 2100 km, whereas values of $\varsigma$ and $L$ vary between 660 and
1460 km. The structures for $b_t^2$ yield Batchelor integral scale as $\lambda
\approx 1600$~km, a correlation length as $\varsigma \approx 640$~km, and an
$e$-folding length as $L \approx 620$~km. The averaged (over $\lambda,
\varsigma$, and $L$) characteristic scale for the $u^2$-fluctuations is
1260$\pm$500~km, and that for $b_t^2$ is 950$\pm$560~km.

The above results show that the characteristic length derived from squared
velocity fluctuations exceeds that derived from the squared
transverse magnetic field fluctuations by 20 - 70\%. 

By analyzing the correlation functions of the velocity components $u_x$ and
$u_y$, we found that the normal and parallel correlation functions are
different. For the case of isotropic hydro-dynamical turbulence $B_p(r) >
B_n(r)$, whereas the correlation functions observed here are in the opposite
sense, i.e., $B_p(r) < B_n(r)$ at all observable scales. The discrepancy might
be attributed to the presence of the magnetic field and/or violation of the
isotropy. The parallel correlation function yields a characteristic length for
the velocity fluctuations of 803$\pm$157~km, whereas the characteristic length
derived from the normal correlation function is 1307$\pm$278~km.

As the spatial resolution improves, the correlation function for $b_z^2$ becomes
narrower and yields smaller estimates of the characteristic length. Since the
correlation functions of $b_z^2$ and $b_t^2$ are similar, one might expect that
with improved solar instruments, $b_t^2$ might well produce smaller
estimates for $\lambda_{\bot}$. Thus, we may perhaps regard the 
magnitudes of $\lambda_{\bot}$ reported here as an upper limit only.

Certainly, the study described here warrants further investigation using
better magnetic field data. For example, the velocity and transverse magnetic
field measurements should refer to a CH area, whereas in this study, the
transverse magnetic field fluctuations were obtained for a quiet sun region.
However, we may be obtaining realistic estimates because, as it is shown
in Figure \ref{fig8}, the correlation functions for CH and the quiet sun areas
are rather similar and produce similar estimates of the characteristic length.
It would be also useful to compare the LCT results with those derived from the
tracking of magnetic elements. This task requires high cadence and a long series
of magnetic fields measured inside CHs, and is left for a future study.

We have shown that the characteristic length of the energy-containing structures
in the photosphere lies in the range of 600-2000~km, which is on average an
order of magnitude lower than the values used currently in models
\citep[e.g.,][]{Matthaeus+1999,Dmitruk+2001}. Taking into account that the
nonlinear dissipation terms in the MHD equations (Eq.(1) in \citet{Zank+2012}),
is inversely proportional to the correlation length of energy containing
structures at the base of the corona (see Eq.16 in \citet{Zank+2012}), our
results here play a critical role in determining the effectiveness of the
coronal turbulence transport models in heating the solar corona and hence in
driving the solar wind.

 It is worthy to note that obtained estimates of the averaged transverse
velocity (about 1.2 km s$^{-1}$) and the characteristic length scale for
$u^2$-fluctuations (about 1300 km), combined with the results of
\citet{Rudiger-Kitchatinov-2011} allowed us to evaluate the turbulent magnetic
diffusivity and cross-helicity in the photosphere. Indeed, when the above 
estimates are used in the expression for turbulent magnetic
diffusivity $\eta_T \approx u_{rms} l/3$, it gives us $\eta_T \approx$ 500 $\pm$
200 km$^2$ s$^{-1}$ ($5\times 10^{12}$ cm$^2$ s$^{-1}$). In turn, the
cross-helicity, according to Eqs. 19,24 in \citet{Rudiger-Kitchatinov-2011}, is
in direct proportion with $\eta_T$. This allows us to conclude that the
magnitude of the cross-helicity (1 G km s$^{-1}$) obtained by
\citet{Rudiger-Kitchatinov-2011} with  $\eta_T=1\times 10^{12}$ cm$^2$ s$^{-1}$,
might be at least five times larger. The importance of the cross-helicity for
estimation of internal solar parameters needed for dynamo was emphasized by
\citet{Kuzanyan+2007} and \citet{Pipin+2011-CrossHelicity}. 

It is interesting that the magnitude of the turbulent magnetic diffusivity
obtained here in a CH for the spatial scale of $\sim$1300~km coincides with the
value of $\eta_T$ inferred from the spectrum of the turbulent diffusion
coefficient in a CH on scale of 1000-1300~km (460-560km$^2$ s$^{-1}$), reported
by \citet[][see Fig. 10 there]{Abramenko+2011-Diff}. Thus, our present results
indirectly support the previous inference that the turbulent diffusivity is a
scale-dependent parameter. Additionally, the agreement between two independent
estimates of $\eta_T$ argues for the reliability of the LCT technique.

We thank the anonymous referee, whose careful reading and comprehensive reports
with critical comments significantly improved the paper. We are thankful to Dr.
B. Lites for offering the processed SOT/SP data, to Drs. B. Vazquez, D.
Hathaway, C. Smith for helpful discussion of these results. BBSO team efforts
were supported by NSF (AGS-1146896, and ATM-0847126), NASA (NNX11AO73G), and
AFOSR (FA9550-12-1-0066) grants.

%\bibliographystyle{$HOME/texmf/tex/latex/aastex/apj}
%\bibliography{$HOME/abramenko}

\end{document}